\documentclass[12pt]{article}
\usepackage{pdproc,epsfig} 

\begin{document}

\renewcommand{\thefootnote}{\alph{footnote}}
  
\title{Implications for
Solar Neutrino Oscillations from
Super-Kamiokande and SNO Data}

\author{Michael B. Smy}

\address{ Department of Physics \& Astronomy, 
4182 Frederick Reines Hall,
University of California, Irvine\\
Irvine, California 92697-4575, USA\\
 {\rm E-mail: smy@solar1.ps.uci.edu}}




\abstract{
Super-Kamiokande uses neutrino-electron elastic scattering to
measure the recoil electron spectrum and zenith-angle dependence
of solar $^8$B neutrinos. SNO has measured the $^8$B neutrino--deuteron
charged-current reaction rate.
The elastic scattering rate, spectrum and zenith-angle dependence
in conjunction with the charged-current reaction rate
favors active neutrino oscillations at large mixing angles
by about 3$\sigma$ over
the no-oscillation hypothesis and small mixing angles.
The analysis is independent of the absolute $^8$B and {\it hep}
flux and assumes two-flavor oscillations described by
mixing angle and mass$^2$ difference.
Two allowed regions at large mixing are found.}
   
\normalsize\baselineskip=15pt

\section{Introduction}
A simple two-neutrino oscillation model ($\nu_e\longrightarrow\nu_x$)
is capable of explaining why
all rates measured by various solar neutrino
experiments\cite{homestake,kam3,sage,gallex,fluxpaper,sno}
are smaller than predictions based on the standard solar model\cite{ssm}.
Elastic neutrino-electron scattering is sensitive to all active
neutrino types. However, the cross section for muon- or tau-type
neutrinos is about 5.2 to 6.8 times smaller than for
electron-type neutrinos.
The neutrino--deuteron charged-current reaction on
the other hand only measures the electron-type neutrino flux.
Neutrino flavor oscillation into muon- or tau-type will
suppress the observed charged-current reaction rate more
strongly than the observed elastic scattering rate.
A combined fit to the elastic scattering precision
data of Super-Kamiokande\cite{fluxpaper} and SNO's measurement
of the charged-current reaction rate\cite{sno}
can therefore indicate active neutrino oscillations.
Since both experiments measure the solar neutrino flux from
the $^8$B branch of the solar neutrino spectrum, the results of such
a combined fit are independent of solar neutrino flux
predictions by the standard solar model.
In addition to this comparison between the rates of Super-Kamiokande
and SNO, the shape of Super-Kamiokande's
recoil electron spectrum and its data on the zenith-angle dependence
of the solar neutrino flux constrain\cite{oscpaper}
the mixing angle $\theta$ and mass$^2$ difference $\Delta m^2$
between the two neutrinos.

The solar neutrino rate measurements
also define allowed areas for these parameters.
The mixing angle of the large mixing angle solution is close to maximal
(around $\sin^22\theta\approx0.8$) and has a range in $\Delta m^2$ of
$8\cdot10^{-6}$eV$^2$ to $3\cdot10^{-4}$eV$^2$. The mixing angle
of the small mixing angle solution is between $2\cdot 10^{-3}$ and $10^{-2}$
for a $\Delta m^2$ between $4\cdot10^{-6}$eV$^2$ and $10^{-5}$eV$^2$.
The low solution is at maximal mixing between $6\cdot10^{-8}$eV$^2$ and
$2\cdot10^{-7}$eV$^2$. Vacuum solutions occur between 
$6\cdot10^{-11}$eV$^2$ and $10^{-10}$eV$^2$ for mixing angles
above $\sin^22\theta>0.5$ and some quasi vacuum
solutions (with lower probability)
between $2\cdot10^{-10}$eV$^2$ and $7\cdot10^{-10}$eV$^2$ at maximum mixing.


\section{Analysis Method}

The oscillation analysis is similar to the analysis of the Super-Kamiokande
zenith angle spectrum~\cite{oscpaper}.
For each energy bin $i$, we form a zenith angle rate difference vector
$\overrightarrow{\Delta_i}$. Its seven zenith components $\Delta_{i,z}$ are
\[
\Delta_{i,z}(\sin^22\theta,\Delta m^2;\alpha,r_{\mbox{\it hep}})=
\frac{\phi^{\mbox{\tiny meas}}_{i,z}}{\phi^{\mbox{\tiny SSM}}_i}-
\alpha\times f\left(E_i,\delta_{\mbox{\tiny corr}}\right)\times
\frac{\phi^{\mbox{\tiny osc}}_{i,z}(\sin^22\theta,\Delta m^2;r_{\mbox{\it hep}})}{\phi^{\mbox{\tiny SSM}
}_i}
\]
where $\theta$ is the two-neutrino mixing angle and $\Delta m^2$
the mass$^2$ difference between the neutrinos. The parameter $\alpha$
normalizes the neutrino flux and $r_{\mbox{\it hep}}$ describes
the {\it hep} flux relative to the $^8$B flux.
The observed rate in
energy bin $i$ and zenith angle bin $z$ is denoted as
$\phi_{i,z}^{\mbox{\tiny meas}}$;
$\phi^{\mbox{\tiny SSM}}_i$ and
$\phi^{\mbox{\tiny osc}}_{i,z}$ (calculated as in\cite{oscpaper})
are the expected event rates in that bin without and with neutrino
oscillation. 
The spectral distortion $f$ due to the correlated
systematic error of $\phi_{i,z}$ is scaled by the parameter
$\delta_{\mbox{\tiny corr}}$.
Similarly, we define the charged-current rate difference
\[
\Delta_{\mbox{\tiny CC}}(\sin^22\theta,\Delta m^2;\alpha,r_{\mbox{\it hep}})=
\frac{\phi^{\mbox{\tiny meas}}_{\mbox{\tiny CC}}}
     {\phi^{\mbox{\tiny SSM}}_{\mbox{\tiny CC}}}-
\alpha\times
\frac{\phi^{\mbox{\tiny osc}}_{\mbox{\tiny CC}}(\sin^22\theta,\Delta m^2;r_{\mbox{\it hep}})}{\phi^{\mbox{\tiny SSM}}_{\mbox{\tiny CC}}}.
\]
To compute $\phi^{\mbox{\tiny osc}}_{\mbox{\tiny CC}}$
the neutrino--deuteron charged-current
cross section calculation
of Ying, Haxton and Henley\cite{deutcross} is used. The oscillation
probability is calculated in a similar way as for
$\phi^{\mbox{\tiny osc}}_{i,z}$.
The $\chi^2$ is defined as
\begin{equation}
\chi^2(\sin^22\theta,\Delta m^2)=\min_{\alpha,r_{\mbox{\it\tiny hep}}}\left(
\sum_{i=1}^{8}
\overrightarrow{\Delta_i}\cdot V_i^{-1}\cdot\overrightarrow{\Delta_i}
+\left(
\frac{\delta_{\mbox{\tiny corr}}}{\sigma_{\mbox{\tiny corr}}}
\right)^2
+\left(
\frac{\Delta_{\mbox{\tiny CC}}}{\sigma_{\mbox{\tiny CC--ES}}}
\right)^2
\right)
\end{equation}
Each energy bin $i$ has a separate $7\times7$
error matrix $V_i$ describing the energy-uncorrelated uncertainty.
$V_i$ is the sum of the statistical error matrix and the
energy-uncorrelated systematic error matrix,
the latter of which is
constructed
assuming full correlation in zenith angle.
The flux normalization factor
$\alpha$ is constrained by the last term containing
the uncertainty
$\sigma_{\mbox{\tiny CC--ES}}$. This uncertainty is the quadratic sum
of SNO's statistical and systematic uncertainty and the
systematic uncertainty of Super-Kamiokande's total rate.
It is the systematic uncertainty of the SK--SNO rate difference
combined with the SNO statistical uncertainty.

\begin{figure}[p]
\centerline{
\epsfxsize 2.7in\epsffile{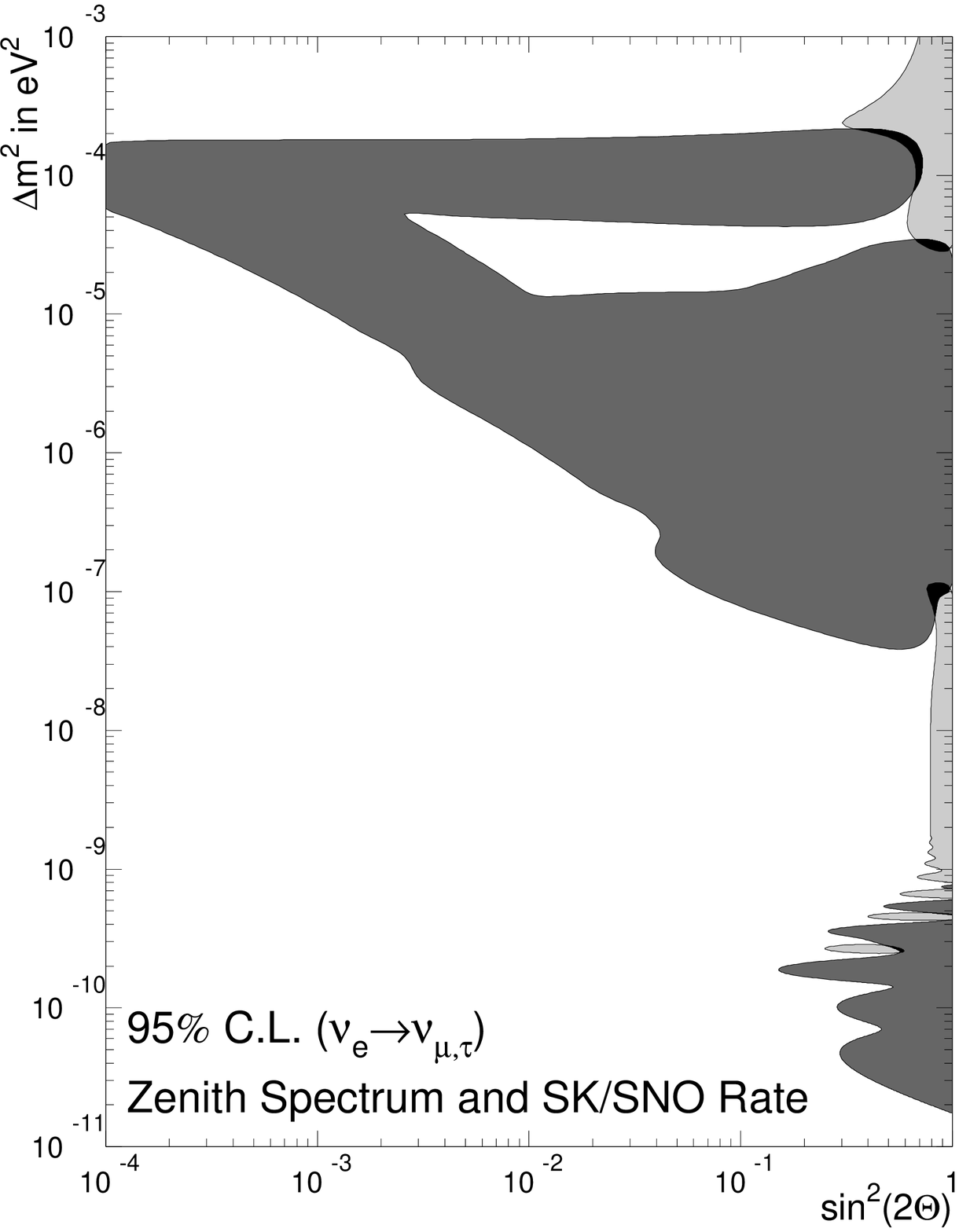}
\epsfxsize 2.7in\epsffile{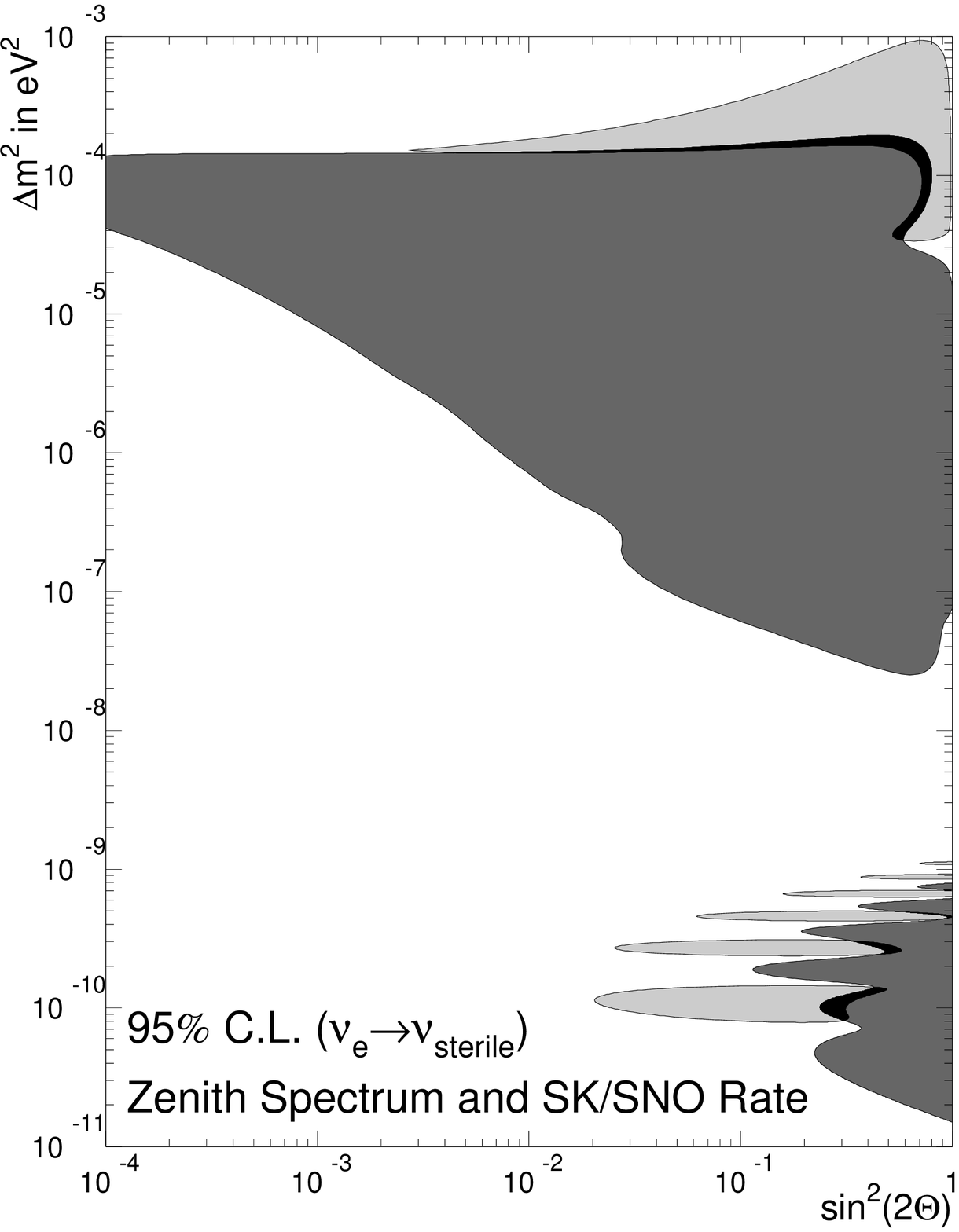}}
\vspace*{-0.15in}
\caption{Excluded regions
(dark shaded area; 95\% C.L.)
using the shape of the zenith angle spectrum\protect\cite{oscpaper}
and allowed regions (light shaded area; 95\% C.L.) using the
rate and zenith angle spectrum and
the charged-current reaction rate
for fully active (left)
and fully sterile (right) two-neutrino oscillations.}
\label{figsno:area}
\vspace*{-0.15in}
\centerline{
\epsfxsize 2.7in\epsffile{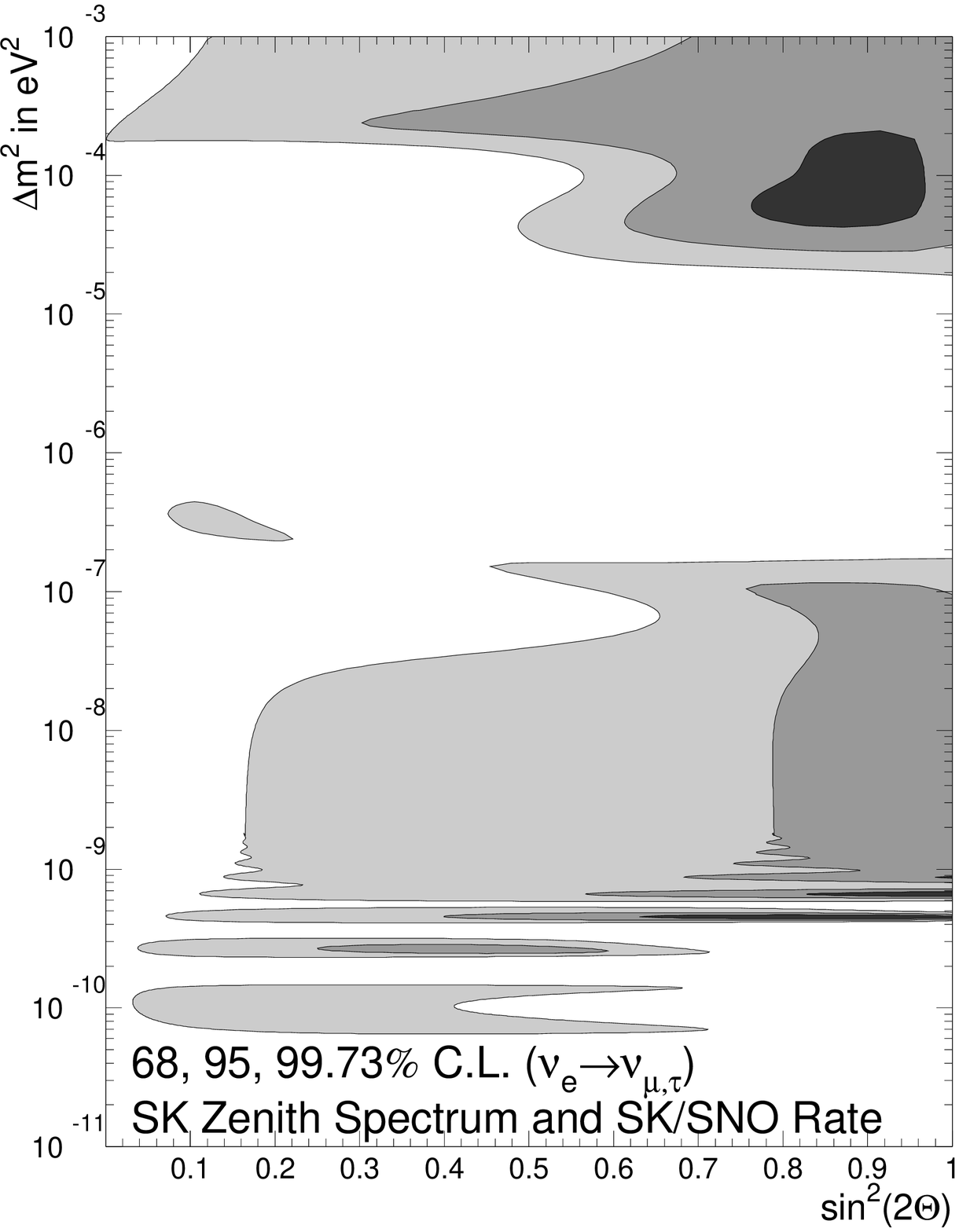}
\epsfxsize 2.7in\epsffile{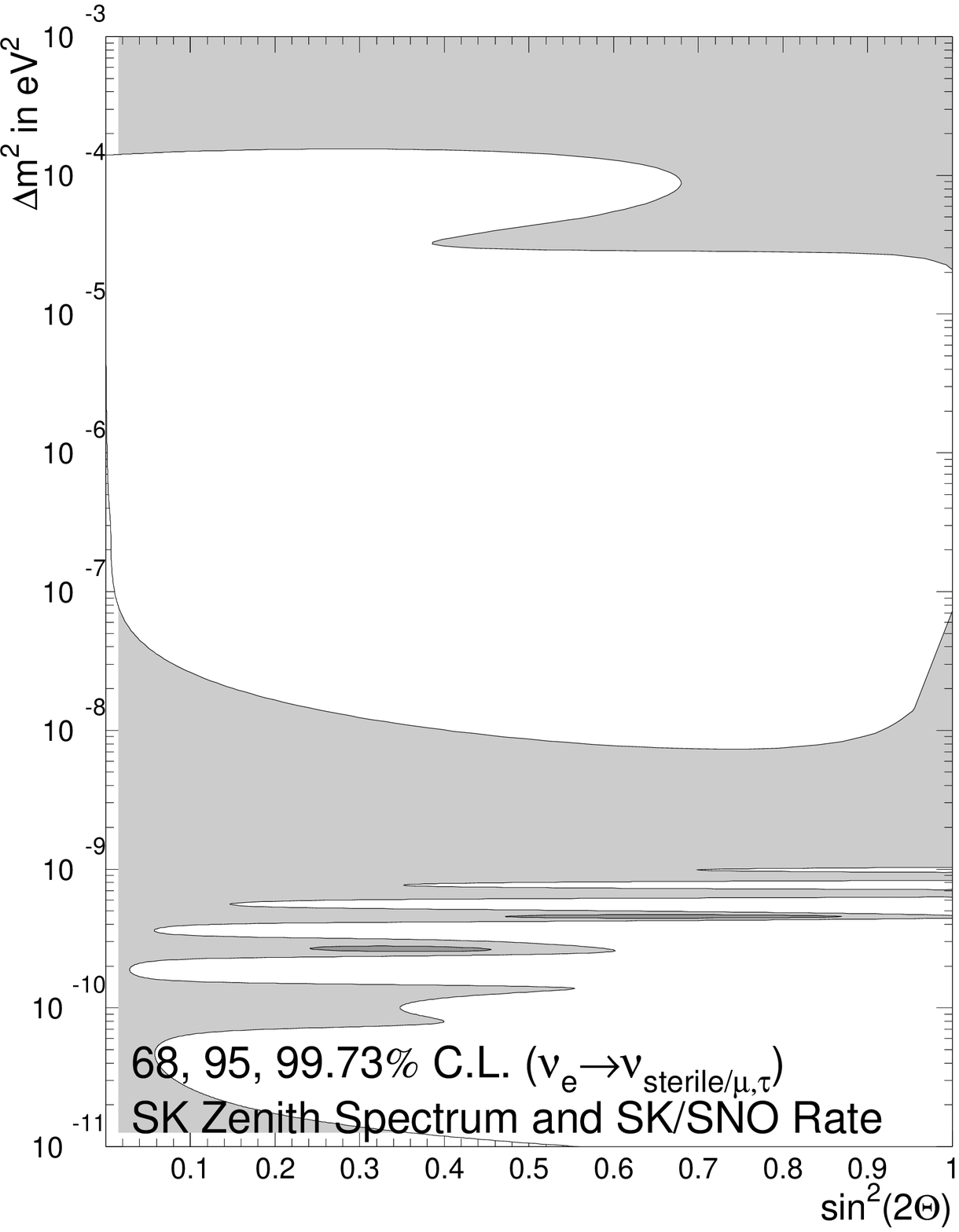}}
\vspace*{-0.15in}
\caption{Allowed regions using the shape of the
zenith angle spectrum and allowed
regions at the one (dark shaded area), two
(shaded area) and three (light shaded area)
sigma level using the
rate and zenith angle spectrum and
the charged-current reaction rate
for fully active (two neutrinos; left)
and fully sterile (three neutrinos; right) oscillations.
The horizontal scale is linear.}
\vspace*{-0.15in}
\label{figsno:threesig}
\end{figure}

\section{Results of the $\chi^2$ Analysis}
Active two-neutrino oscillations are favored by $3\sigma$
over the no-oscillation hypothesis. The allowed parameter
regions are shown in figure~\ref{figsno:area} (logarithmic
scale for sin$^22\theta$) and
figure~\ref{figsno:threesig} (linear scale).
The best fit is at $\Delta m^2=6.6\cdot10^{-10}$eV$^2$
and maximal mixing. The $\chi^2$ is 37.9 with 41
degrees of freedom (44 variables from the
Super-Kamiokande zenith angle spectrum, one from the
SNO rate; four parameters are minimized).
The best-fit $^8$B flux is 78\% of the standard solar model prediction,
the best-fit {\it hep} flux is zero.
The allowed areas in figure~\ref{figsno:area} for
the sterile case are based on a two-neutrino oscillation
analysis with two parameters ($\sin^22\theta$ and $\Delta m^2$)
and a minimum in the
sterile vacuum region. This $\chi^2$ minimum is
worse by 6.4 compared to the best-fit active solution.
A one parameter (sterile content of the
oscillation regardless of $\sin^22\theta$ and $\Delta m^2$)
interpretation of this $\chi^2$ difference implies that
fully sterile oscillations (sterile content=1) are
disfavored by more than $2.5\sigma$ compared with
fully active two-neutrino oscillations (sterile content=0).
The sterile case of figure~\ref{figsno:threesig}
uses the sterile content of the oscillation as
a third parameter and the best-fit active oscillation
solution as the minimum.
The small mixing angle
solution is disfavored in both the fully active
and the fully sterile case.

\begin{figure}[p]
\centerline{
\epsfxsize 3.0in\epsffile{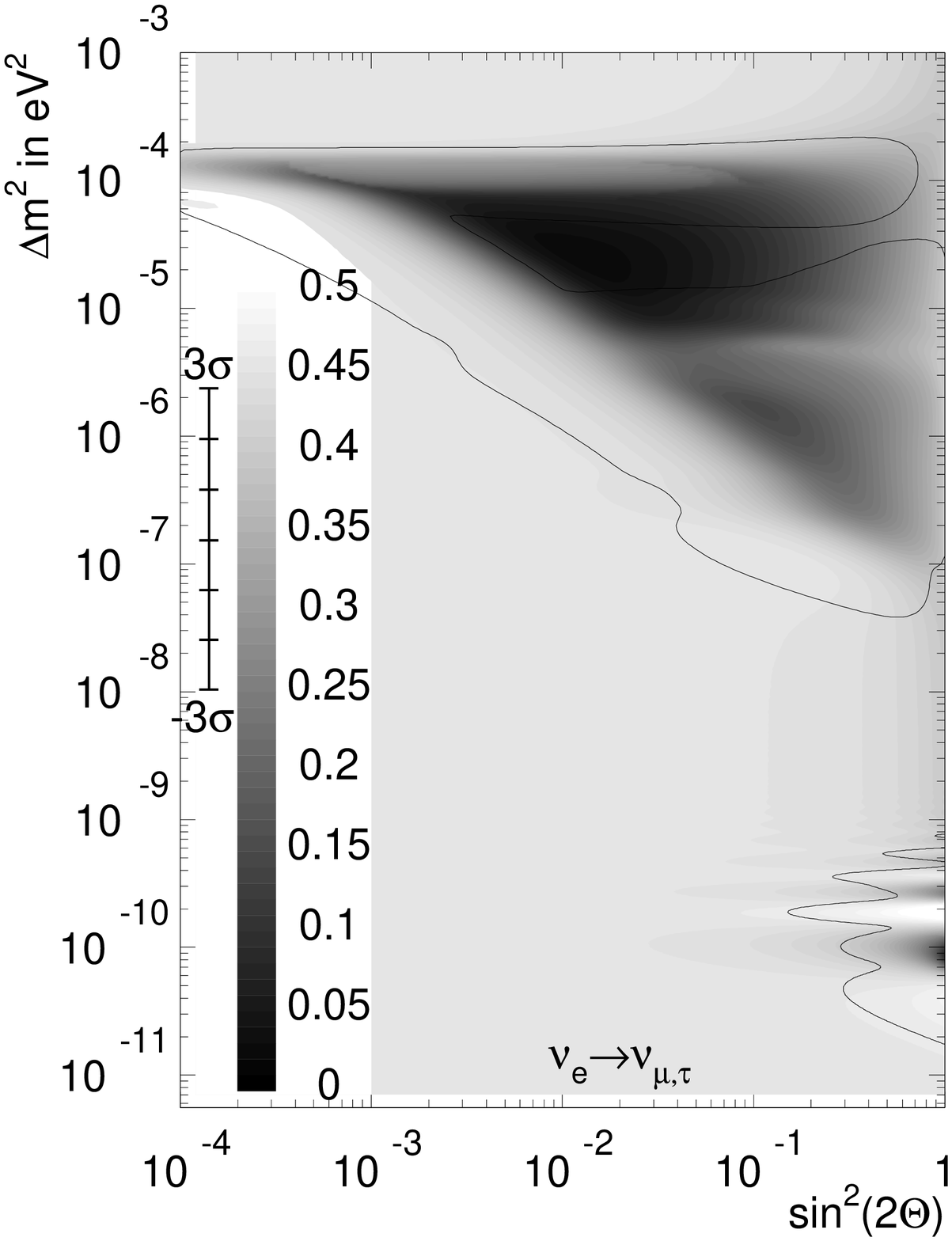}
\epsfxsize 3.0in\epsffile{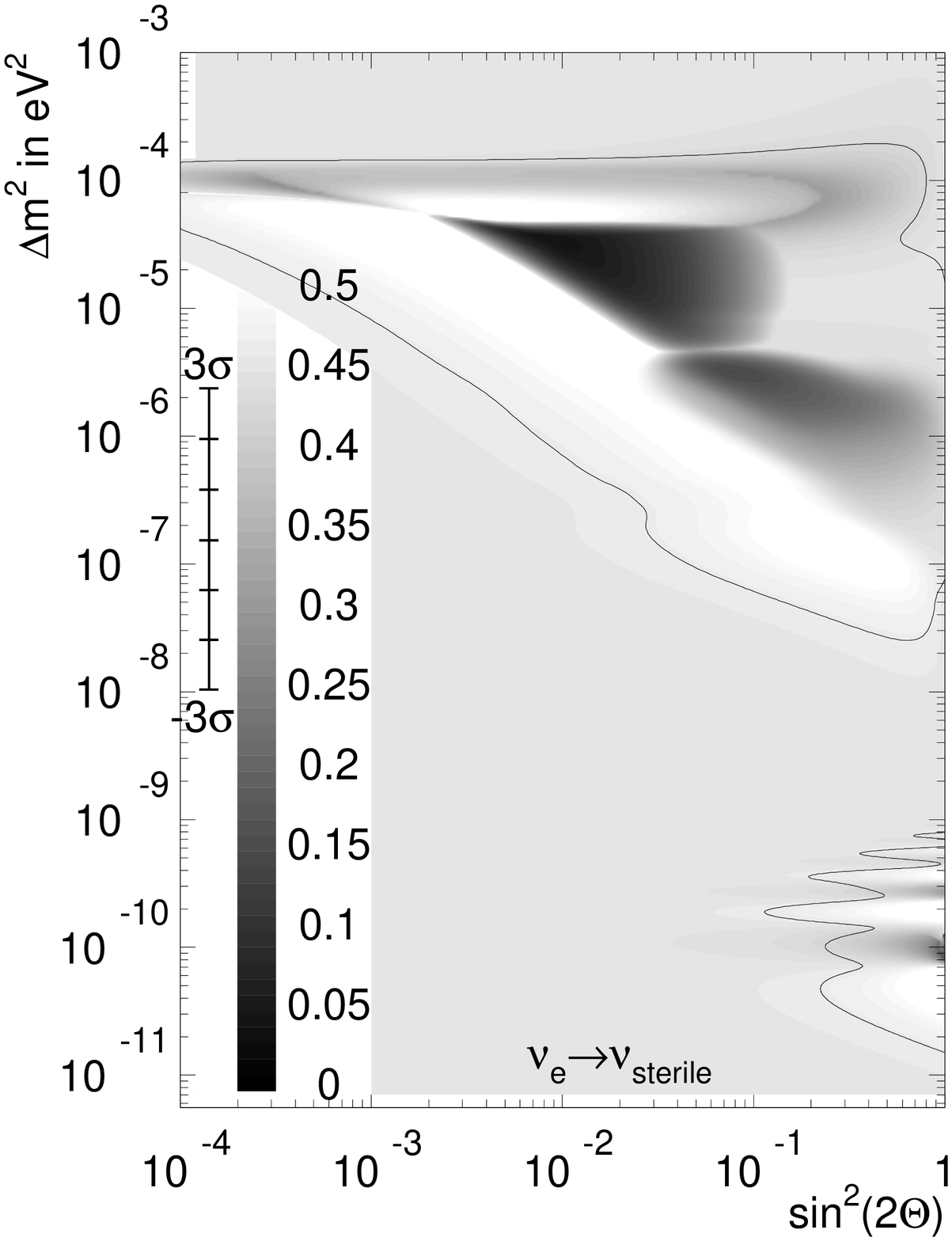}}
\vspace*{-0.15in}
\caption{Best-fit charged-current rate in SNO as a function
of mixing and mass$^2$ difference for fully active
(left) and fully sterile (right) two-neutrino oscillations.
The black lines enclose the area excluded by the
shape of the Super-Kamiokande zenith angle spectrum\protect\cite{oscpaper}
at 95\% C.L.}
\label{fig:snocc}
\vspace*{-0.15in}
\centerline{
\epsfxsize 3.0in\epsffile{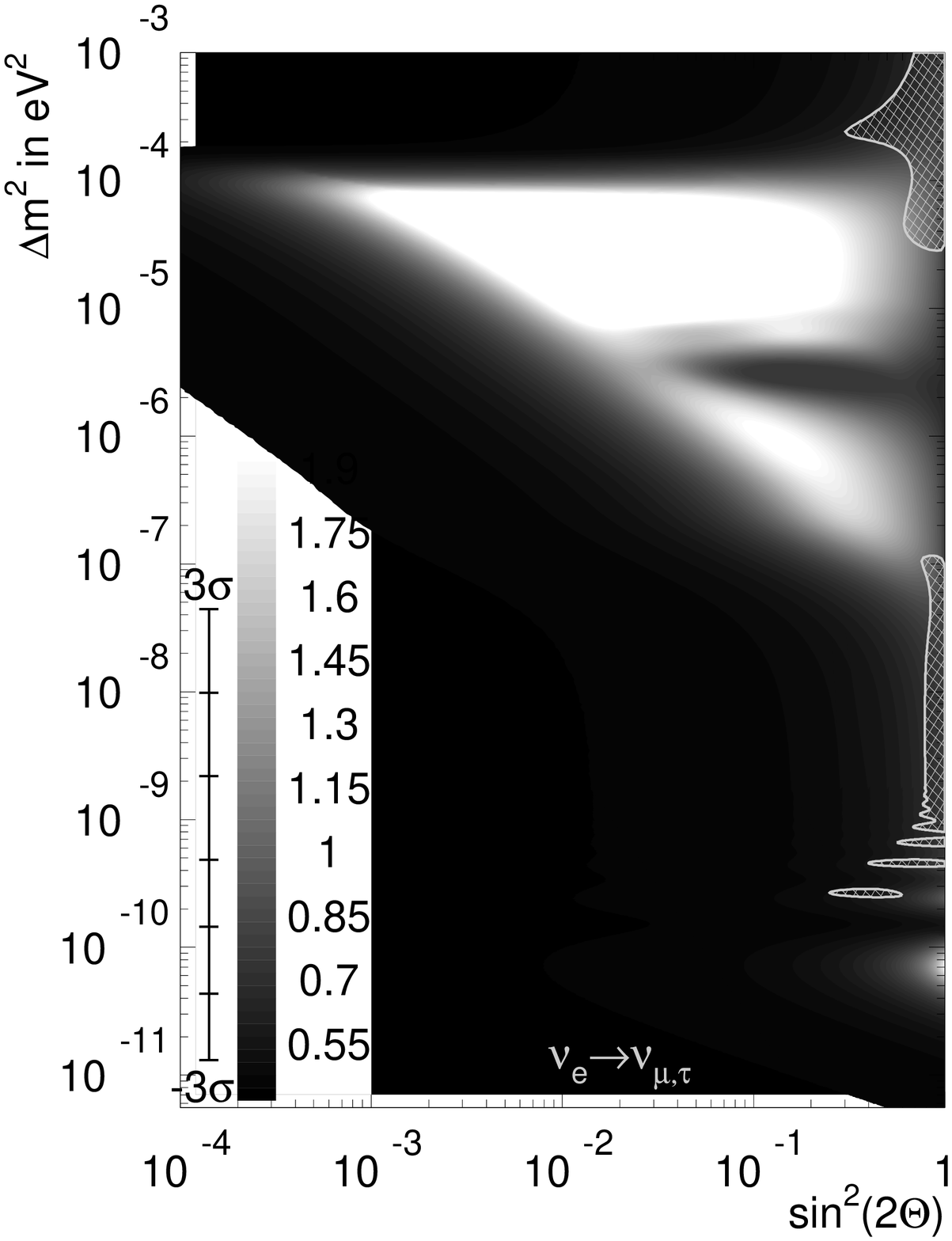}
\epsfxsize 3.0in\epsffile{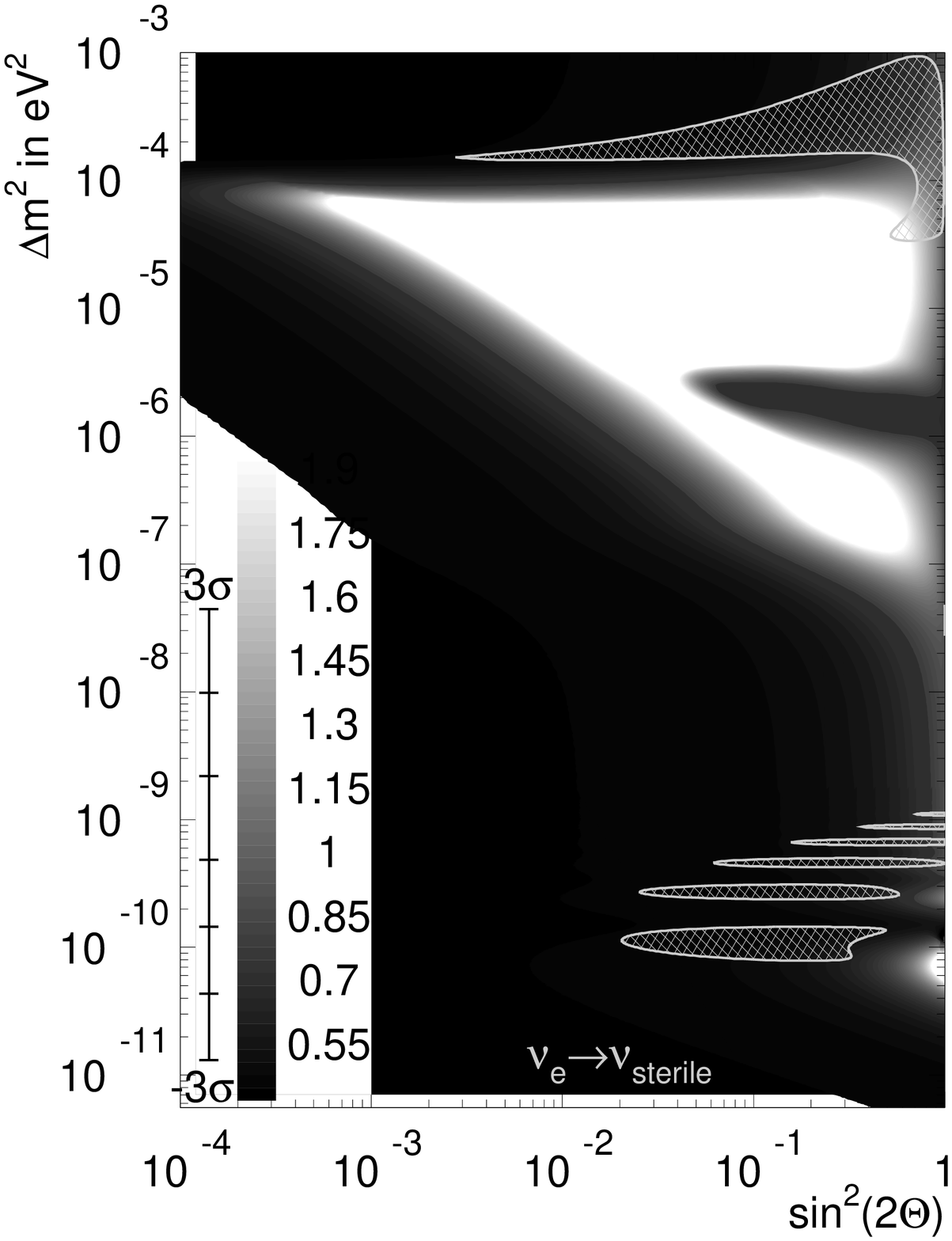}}
\vspace*{-0.15in}
\caption{Best-fit $^8B$ flux in units of the standard
solar model prediction as a function
of mixing and mass$^2$ difference for fully active
(left) and fully sterile (right) two-neutrino oscillations.
The hatched areas are allowed at 95\% C.L. by the Super-Kamiokande
zenith angle spectrum and the Super-Kamiokande--SNO rate comparison.}
\label{fig:b8}
\vspace*{-0.15in}
\end{figure}

Figure~\ref{fig:snocc} shows the best-fit charged-current
rate (to be compared directly with the SNO measurement).
Figure~\ref{fig:b8} displays the best-fit $^8$B flux
in units of the standard solar model prediction. In case
of the active large mixing angle and low solutions,
the best-fit $^8$B agrees with the prediction within the
theoretical uncertainty. Only some active quasi-vacuum solutions
disagree by more than $2\sigma$ from the prediction.
The allowed areas shown in the sterile case use
the best-fit sterile solution  (disfavored by $>2.5\sigma$)
as the minimum.
In general, agreement of the best-fit $^8$B flux with the prediction
is poor in those areas. In some parts of a `sterile large mixing angle
solution', however, the $^8$B flux agrees well.

Allowed areas
are near the active large mixing angle solution and a band
at maximum mixing
extending from the low solution down to the quasi-vacuum
region. Figure~\ref{fig:snocc} shows the reason why the small
mixing angle solutions and most of the vacuum solutions
are disfavored. The best-fit charged-current rate is in agreement with
the SNO measurement mostly
in those areas that disagree with the Super-Kamiokande
zenith angle spectrum. In the active case,
only the upper part of the large mixing
angle solution and a small band extending from the low
to the quasi vacuum solution are outside the Super-Kamiokande
excluded area {\it and} lead to an acceptable fit of the
SNO rate. 

\begin{figure}[p]
\centerline{
\epsfxsize 3.1in\epsffile{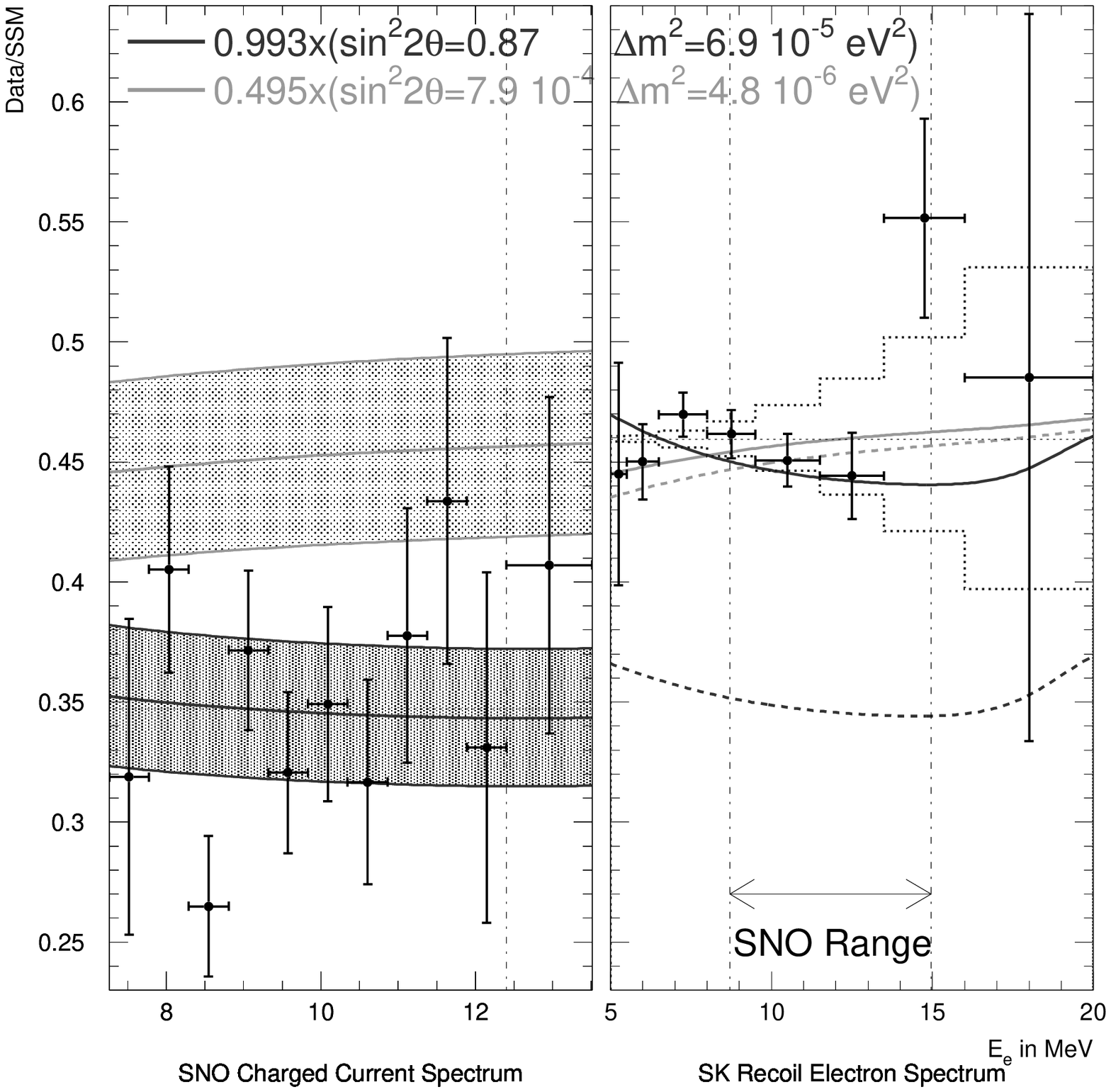}
\epsfxsize 3.1in\epsffile{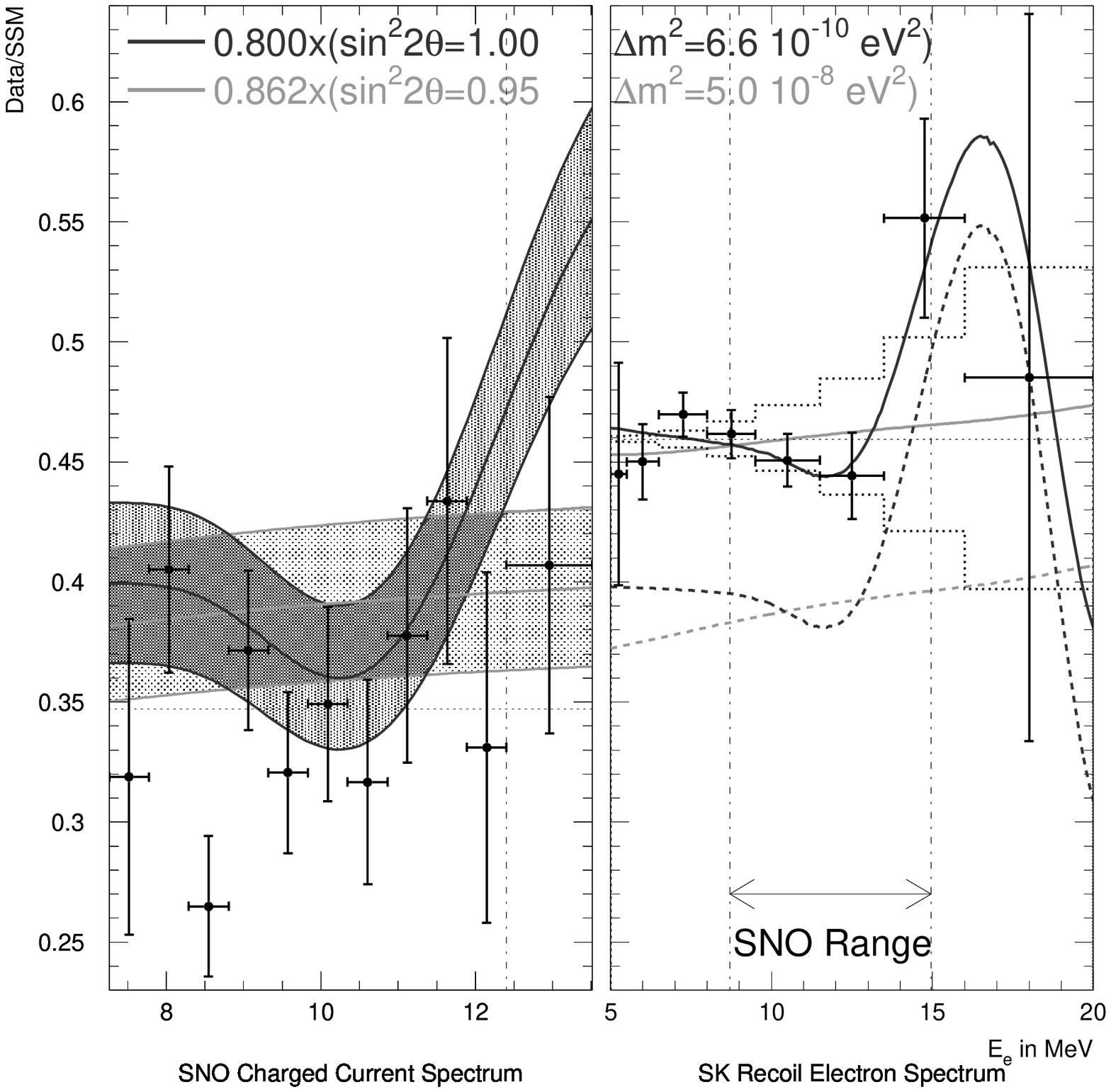}}
\caption{SNO Spectrum (left panel) and
SK Spectrum (right panel) compared with the
best-fit large mixing angle (left figure; dark grey), small mixing
angle (left figure; light grey), low (right figure; light grey)
and vacuum (right figure; dark grey) solution predictions.
The bands around the SNO predictions indicate
the systematic uncertainty for the SK---SNO
flux difference. The dashed lines below
the SK predictions are the contributions due
to electron-type neutrinos only. The dotted
histograms show the $\pm1\sigma$ systematic
energy-correlated uncertainty.}
\label{fig:spec}
\centerline{
\epsfxsize 3.1in\epsffile{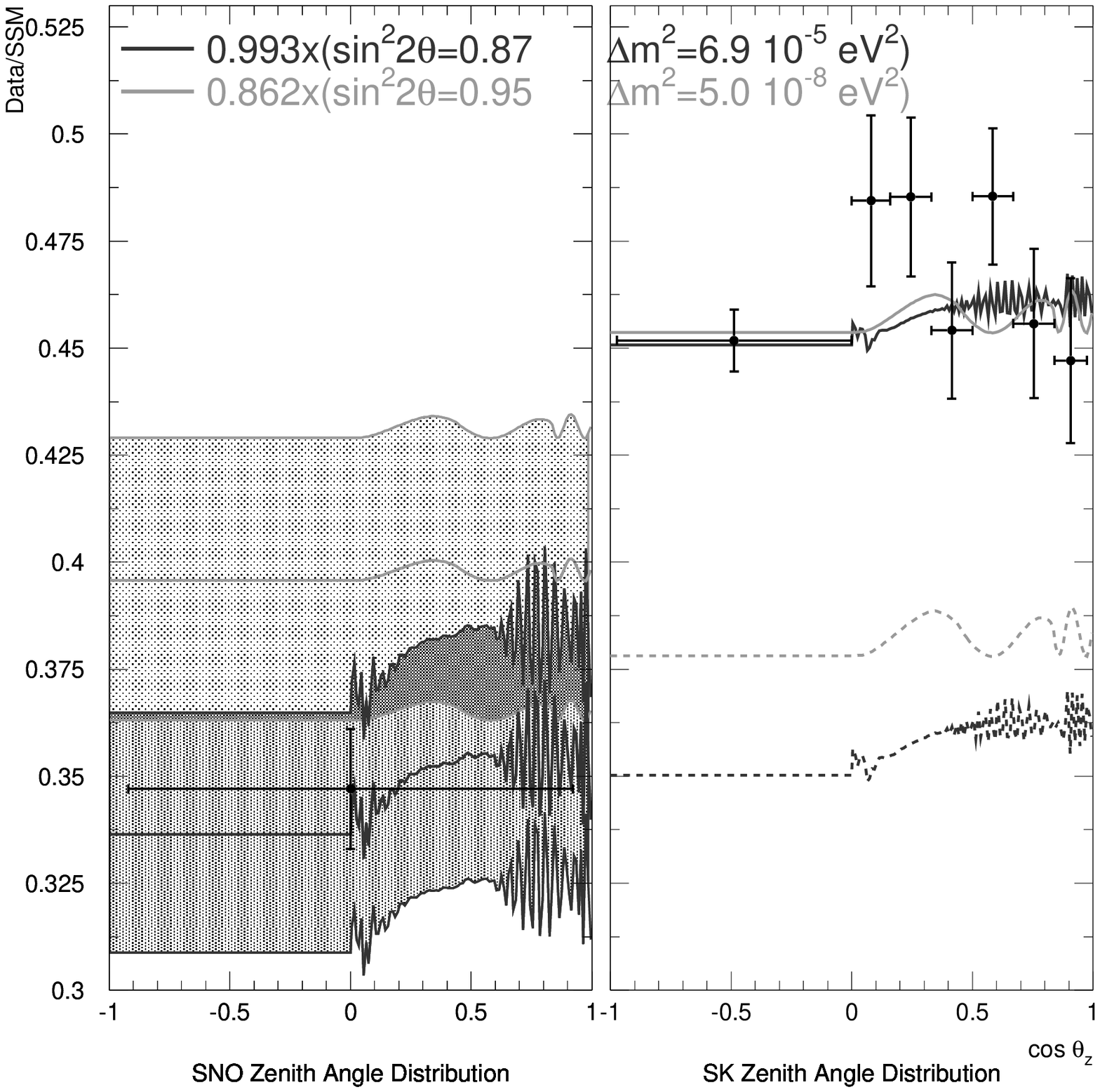}}
\caption{Zenith Angle Distribution for
SNO (left panel) and SK (right panel)
compared with the best-fit large mixing
angle (dark grey) and 
low solution (light grey) predictions.}
\label{fig:dn}
\end{figure}

In the sterile case, no good agreement is found outside
the Super-Kamiokande excluded area.
In particular, active small mixing angles are disfavored
since they either show spectral distortions and core enhancement 
(and disagree with the Super-Kamiokande spectrum) or
have large electron-type survival probabilities. 
In the latter case, the flux difference between 
Super-Kamiokande and SNO cannot be explained.
The best fit close to the small mixing angle solution
(see light grey lines and band on the left-hand side of figure~\ref{fig:spec})
has almost no spectral distortion and a
survival probability around 90\%. This is due to the
very small mixing angle
($\sin^22\theta<10^{-3}$). The contribution
of muon- or tau-type neutrinos to the SK rate (see
difference between dashed and solid line) is very small;
the SK and SNO rates are predicted to be very similar. Since the
survival probability is so large, the $^8$B flux is only
about 50\% of the standard solar model prediction.

The best-fit large mixing angle solution (dark grey lines and
band in the left-hand side of figure~\ref{fig:spec})
on the other hand can
reconcile the flat spectrum and a low survival probability (around
35\%). Large mixing angle solutions
with strong day-night asymmetries are disallowed
by the Super-Kamiokande zenith angle distribution. At present, SNO data does
not have enough precision to observe the weaker day-night
asymmetry predicted by the best-fit large mixing angle solution
(see dark grey lines and band in figure~\ref{fig:dn}).
The $^8$B flux agrees with the standard solar model prediction.
The best-fit {\it hep} flux is about 2.7 times larger than predicted. Note
that all displayed spectrum predictions in figures~\ref{fig:spec}
and~\ref{fig:dn} assume the
standard solar model {\it hep} flux.

The quasi vacuum solution (dark grey lines and
band in the right-hand side of figure~\ref{fig:spec})
follows best the shape of the spectrum. The survival probability
of about 47\% is just small enough to accommodate the SK--SNO
rate difference within systematic uncertainty, assuming the
$^8$B flux is about $1.2\sigma$ below the standard solar model
value. The best-fit {\it hep} flux is zero.

The spectrum of the low solution (light grey lines and
band in the right-hand side of figure~\ref{fig:spec})
fits somewhat worse than the spectrum of the large mixing angle
solution. It is also disfavored by its zenith angle variation
(see light grey lines and band in figure~\ref{fig:dn}).
A survival probability of about 45\% accommodates the
SK--SNO rate difference within systematic uncertainty. The $^8$B
flux is about 1 standard deviation below the standard
solar model prediction, the best-fit {\it hep} flux is twice
as large as in the standard solar model.

\section{Conclusions}
Combining Super-Kamiokande and SNO solar neutrino data places strong
constraints on solar neutrino oscillations
independently of standard solar model neutrino flux
predictions. Large mixing angle
active oscillation solutions are favored by about $3\sigma$ over
small mixing angles. Fully sterile oscillations are disfavored
by more than $2.5\sigma$. Two
allowed areas (one above $\Delta m^2=3\cdot10^{-5}$eV$^2$ and
one below $\Delta m^2=10^{-7}$eV$^2$) are found. For most active
solutions, the best-fit $^8$B flux agrees with the standard solar model
prediction within the given uncertainty. For most sterile solution,
the best-fit $^8$B flux does not agree well with the standard solar model
prediction. The electron-type survival
probability for $^8$B neutrinos is expected to be between 30\% and 47\%.
Spectral distortions and zenith angle variations should also be small
for $^8$B neutrinos.

\newpage
\section{Acknowledgements}
The author acknowledges the cooperation of the Kamioka Mining and
Smelting Company.  The Super-Kamiokande detector has been built and
operated from funding by the Japanese Ministry of Education, Culture,
Sports, Science and Technology, the U.S. Department of Energy, and the
U.S. National Science Foundation.

\end{document}